\pdfoutput=0

\newcommand{\msun}{M$_{\odot}$}
\newcommand{\hii}{HII }
\newcommand{\am}{NH$_{3}$}
\newcommand{\cyano}{HC$_3$N}

\newcommand{\meth}{CH$_3$OH}
\newcommand{\methcy}{CH$_3$CN}

\newcommand{\cm}{cm$^{-3}$}

\newcommand{\kms}{~km~s$^{-1}$\,}
\def\degr{^{\circ}}

\documentclass{iau}
\usepackage{graphicx}

\title[JD 11.~~Radio Molecules in the GC] 
{A Radio Survey of Galactic Center Clouds}

\author[E.A.C. Mills et al.]   
{E.A.C. Mills$^{1}$, C.C. Lang$^{2}$, M.R. Morris$^{3}$, J. Ott $^{1}$,
N. Butterfield$^{2}$, D. Ludovici$^{2}$, S. Schmitz$^{2}$
 \and A. Schmiedeke$^{4}$}

\affiliation{$^1$National Radio Astronomy Observatory (email: {\tt bmills@aoc.nrao.edu})
$^2$Dept. of Physics \& Astronomy, University of Iowa
$^3$Dept. of Physics \& Astronomy, University of California-Los Angeles 
$^4$I. Physikalisches Institut, Universit\"{a}t zu K\"{o}ln}

\pubyear{2014}
\volume{303} 
\pagerange{119--126}
\jname{Radio Molecules in Galactic Center Clouds }
\editors{A.C. Editor, B.D. Editor \& C.E. Editor, eds.}
\begin{document}

\maketitle

\begin{abstract}

We present a radio survey of molecules in a sample of Galactic center molecular clouds, including M0.25+0.01, the clouds near Sgr A, and Sgr B2. The molecules detected are primarily \am\, and \cyano; in Sgr B2-N we also detect nonmetastable \am, vibrationally-excited \cyano, torsionally-excited \meth, and numerous isotopologues of these species. 36 GHz Class I  \meth\, masers are ubiquitous in these fields, and in several cases are associated with new \am\,\hspace{-0.15cm} (3,3) maser candidates. We also find that \am\, and \cyano\, are depleted or absent toward several of the highest dust column density peaks identified in submillimeter observations, which are associated with water masers and are thus likely in the early stages of star formation.


\end{abstract}

The central 300 parsecs of the Galaxy contain one of the largest reservoirs of molecular gas in the Galaxy. Thus far, however, the large-scale distribution and kinematics of this molecular gas have only been probed at arcminute ($\sim$ 2-3 pc) resolutions \cite[(e.g., Bally et al. 1987, Jones et al. 2013)]{Bally87,Jones13}. This survey is a first step toward a uniform study of Galactic center gas on sub-parsec scales. Ultimately, these survey data will probe the temperature, density, and kinematics of a sample of clouds at 2-3$''$ ($\sim$0.1 pc) resolution. 

 \vspace*{-0.4 cm}
\section{Observations and Data Calibration}
\label{obs}
Observations were made using the new WIDAR correlator in the hybrid DnC array of the Karl G. Jansky Very Large Array (VLA). Ka-band data (27-36 GHz) were observed on 7 and 13 January 2012, and K-band data (24-25 GHz) on 8 and 14 January 2012.  The data consist of 18 pointings in 6 clouds ( Sgr B2 M\&N, M0.25+0.01, M-0.11-0.08, M-0.02-0.07, the CND, and M-0.13-0.08). 

The survey was designed to cover a large number of \am\, transitions, which trace gas with densities  $\gtrsim10^3$ \cm, and can be used to measure kinetic temperature.  Eight transitions of \am\, with energies from 20 to 840 K above the ground state are observed, which are collectively sensitive to a wide range of gas temperatures. In addition, the observations cover multiple transitions of \cyano\, and \meth. 

The observations in each band (K and Ka) are divided into two separate, continuous subbands 0.86 GHz wide, each comprised of 7 spectral windows. The typical spectral resolution is 250 kHz ($\sim3$ \kms), however for three spectral windows, covering (1) the \am\, (1,1) and (2,2) lines and their hyperfine structure, (2) the 36.1 GHz \meth\, maser line, and (3) the \methcy\, (2$_k$-1$_k$) transitions, the resolution is doubled to better resolve the line structure. 

The phase calibrator (J1744-3116) was observed approximately every 15 minutes. The bandpass calibrator was J1733-1304. The data were flux calibrated using 3C286, observed at elevations comparable to the Galactic center sources ($\sim 15\degr-30 \degr$).  

 After calibration with the {\it CASA} reduction package, the data were imaged using the CLEAN algorithm in {\it CASA} and a CLEAN cycle of several thousand iterations, resulting in noise in the final images which was typically $\sim$1 mJy beam$^{-1}$ channel$^{-1}$ for the spectral line images. The full width at half maximum (FWHM) of the synthesized clean beam ranged from $2''.3 \times 2''.5$ at 25 GHz to $1''.9 \times 2''.2$ at 27 GHz and $1''.5 \times 1''.7$ at 36 GHz. 
 
\vspace*{-0.2 cm}
\section{Results}
Below, we present preliminary images of M0.25+0.01, clouds near Sgr A, and Sgr B2.  

\begin{figure}[t]
 \vspace*{-0.2 cm}
\begin{center}
 \includegraphics[width=5in]{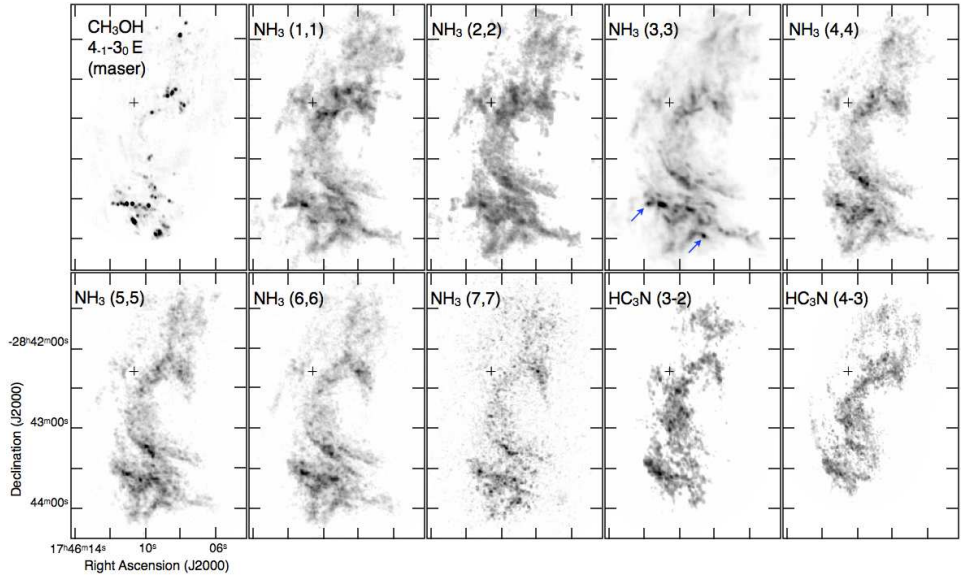} 
\vspace*{-0.1 cm}
 \caption{Peak intensity maps of molecular lines in M0.25+0.01, including 36 GHz methanol masers, 7 transitions of \am, and two transitions of \cyano. The cross indicates the location of an H$_2$O maser \cite[(Lis et al. 1994)]{Lis94} coinciding with a dust continuum peak \cite[(Kauffmann et al. 2013)]{Kauffmann1}. Examples of candidate \am\, (3,3) masers are indicated with arrows.}
   \label{fig1}
\end{center}
\end{figure}

\vspace{0.2cm}
{\underline{\bf M0.25+0.01}}: This cloud is one of the most massive in the CMZ \cite[(M$\sim1-2\times10^5$ \msun; Lis et al. 1994, Longmore et al. 2012)]{Lis94,Longmore12}. However, unlike other massive CMZ clouds, there is no evidence in this cloud for ongoing star formation apart from a single water maser \cite[(Lis et al. 1994, Longmore et al. 2012, Kauffmann et al. 2013)]{Lis94,Longmore12,Kauffmann13}. 

	    Mapping this cloud, we find the morphology of the \am\, transitions is almost identical (apart from a few candidate masers in the (3,3) lines; Fig. 1): there is no sign of temperature gradients. The (1,1) and (2,2) transitions are both very optically-thick, ($\tau$ up to 4-7), as measured from the ratio of their hyperfine satellites. From the optically-thin (4,4), (5,5), and (7,7) lines, we measure a median temperature of $\sim$100 K.  
	    
	    The distribution of \cyano\, is similar to \am, but is stronger toward the center of the cloud, and may trace denser gas. We also detect numerous 36.1 GHz \meth\, point sources, the distribution of which closely follows the morphology of the \am\, gas (Fig. 1). The majority of the \meth\, sources are relatively weak, with intensities of 0.5 - 1 Jy beam$^{-1}$. However, more than 80\% can be identified as masers, having brightness temperatures greater than the hottest gas identified in this cloud \cite[(400 K; Mills \& Morris 2013)]{Mills13}. Dozens of such candidate masers are found in every cloud surveyed.  As these are Class I masers, they are indicative of shocks, but in the turbulent environment of the Galactic center, they likely do not indicate star formation. 

\vspace{0.2cm}
{\underline{\bf Sgr A}}: There are three main clouds near Sgr A: M-0.02-0.07, M-0.13-0.08, and the circumnuclear disk (CND), which surrounds the supermassive black hole, Sgr A*. In projection, these clouds lie within the central 10 pc of the Galaxy, and are believed to be physically close as well, based upon their interactions with each other and with the gravitational potential of Sgr A* \cite[(Coil \& Ho 2000, Herrnstein \& Ho 2005)]{Coil00,HH05}.  
	     
\begin{figure}[t]
\vspace*{-0.3 cm}
\begin{center}
 \includegraphics[width=5in]{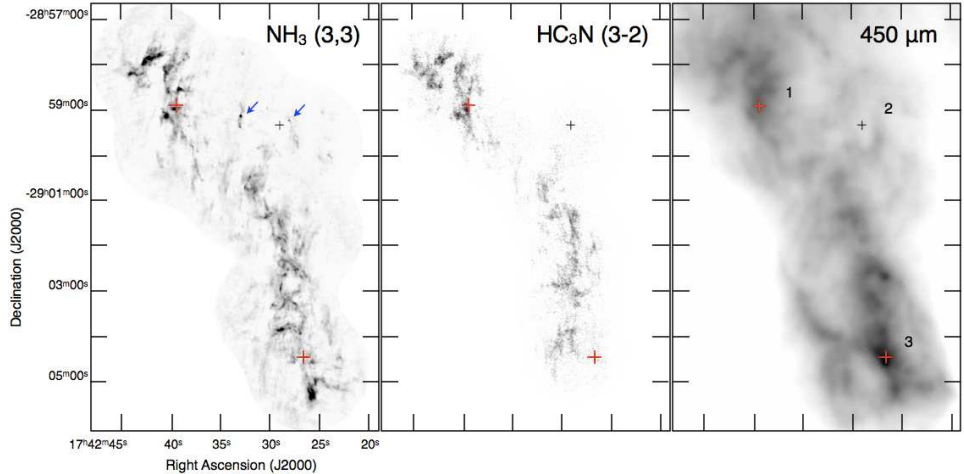} 
\vspace*{-0.1 cm}
 \caption{Peak intensity maps of \am\, (3,3) (Left) and \cyano\, 3-2 (Center) in Sgr A, compared to 450 micron dust continuum \cite[(Right; Pierce-Price et al. 2000)]{PP00}. Crosses indicate dust continuum peaks in M-0.02-0.07 (1), M-0.13-0.08 (3), and the location of the supermassive black hole Sgr A* (2). Examples of candidate \am\, (3,3) masers are indicated with arrows.}
   \label{fig2}
\end{center}
\end{figure}	     	     
	     
The strongest sources of \am\, emission are M-0.02-0.07, and a clump at the southern tip of M-0.13-0.08 (Fig. 2). However, this clump is entirely absent in \cyano. As in M0.25+0.01, there is also no \am\, or \cyano\, detected toward the strongest dust continuum peak in M-0.13-0.08 and its associated H$_2$O masers \cite[(Guesten \& Downes 1983; Sjouwerman et al. 2002)]{GD83,Sjou02}.  
               
               \cyano\, is not detected toward the CND, which is either an indication that gas in this region is less dense, or that \cyano\, is under-abundant or destroyed in this environment \cite[(Mart\'{i}n et al. 2012)]{Martin12}. \am\, emission from the CND is weak in comparison to emission from M-0.02-0.07 and M-0.13-0.08, although several strong, compact sources of \am\, (3,3) emission (likely masers) are found here, coincident with observed 36 and 44 GHz \meth\, masers \cite[(Sjouwerman et al. 2010, Yusef-Zadeh et al. 2008)]{Sjou10,YZ08}. 

\vspace{0.2cm}
{\underline{\bf Sgr B2}}: This is the most massive Galactic center cloud, and hosts extremely active star formation traced by dozens of ultracompact \hii regions \cite[(e.g., De Pree et al. 1998)]{DP98}.

The structure of the cloud, as seen in \am\,\hspace{-0.15cm} (3,3), is a wide filament oriented in the southeast/northwest direction (Fig. 3). \am\,\hspace{-0.15cm} (3,3) emission is detected from the `N' and `M' sub-millimeter cores, as well as likely (3,3) masers in the south (Mart\'{i}n-Pintado et al. 1999). Emission from most other molecules is confined to N, although all three sub-millimeter cores (N,M,S) are detected in J$_2$-J$_1$ E-type \meth\, transitions. In the primary hot core in N (SMA-1, v$\sim$63 \kms), emission from these lines of \meth\, and its $^{13}$C isotopologue, as well as from a torsionally-excited line of A$^+$-type \meth\, (12$_ 2$- 11$_ 1$, v$_T$=1) traces a ringlike structure of diameter 5$''$ (Fig. 3), first mapped in CH$_3$CH$_2$CN (Hollis et al. 2003). A second hot core in N \cite[(SMA-2, v$\sim$75\kms, 5$''$ to the north; Liu \& Snyder 1999, Qin et al. 2011)]{LS99, Q11} is also seen in the J$_2$-J$_1$ lines of \meth\, and $^{13}$CH$_3$OH.

Metastable \am\, emission in SMA-1, as previously detected by Vogel et al. (1987), is extremely optically thick: hyperfine satellites are clearly detected in all lines, including (9,9). Emission from the \am\, satellite lines (which are more optically thin than the main lines) as well as the optically-thin isotopologue $^{15}$\am, peaks on the southern edge of this ring, and is roughly co-spatial with emission from vibrationally-excited transitions of \cyano\, and non-metastable \am. However, the 4$_{14}$-4$_{04}$ transition of singly-deuterated ammonia (NH$_2$D) is not detected toward SMA-1, and is only seen toward SMA-2, suggesting the deuterated fraction of this core may be higher than previously estimated for `N' as a whole (Peng et al. 1993).

\begin{figure}[t]
 \vspace*{-0.2 cm}
\begin{center}
 \includegraphics[width=5in]{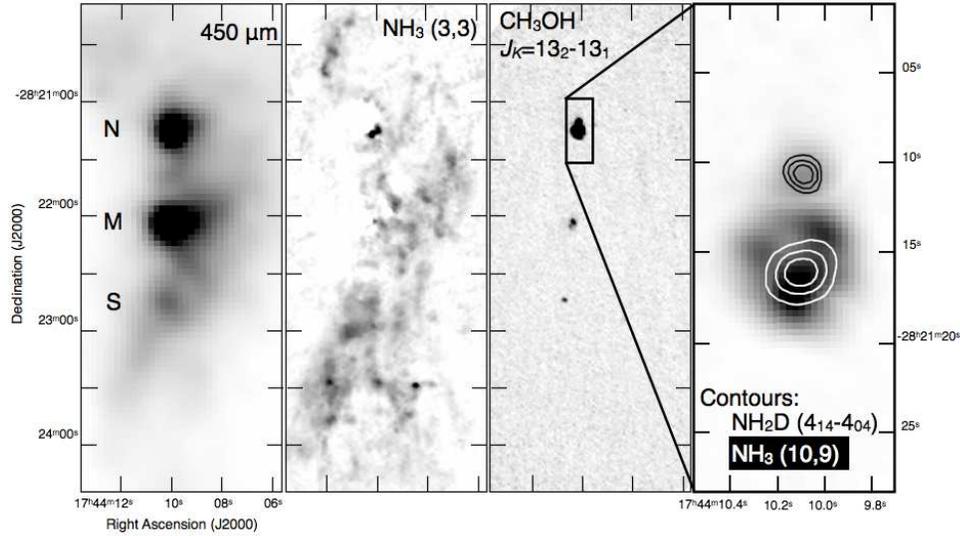} 
 \vspace*{-0.3 cm}
 \caption{Peak intensity maps of \am\, (3,3) (Center) and \meth\, 13$_2$-13$_1$ (Rightmost two panels) in Sgr B2, compared to 450 micron dust continuum \cite[(Left; Pierce-Price et al. 2000)]{PP00}. The far-right panel shows a zoom of Sgr B2-N and the two hot cores: SMA-1(white contours) the most chemically-rich source in this survey and the location of the strongest nonmetastable \am\, emission, and SMA-2 (black contours), where we detect NH$_2$D}
   \label{fig3}
\end{center}
\end{figure}

\end{document}